\documentclass[sigconf]{aamas} 

\usepackage{algorithm}
\usepackage{algorithmic}

\usepackage{balance} 

\doi{NQKD4743}

\makeatletter
\gdef\@copyrightpermission{
  \begin{minipage}{0.2\columnwidth}
   \href{https://creativecommons.org/licenses/by/4.0/}{\includegraphics[width=0.90\textwidth]{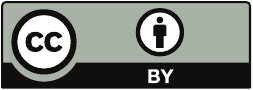}}
  \end{minipage}\hfill
  \begin{minipage}{0.8\columnwidth}
   \href{https://creativecommons.org/licenses/by/4.0/}{This work is licensed under a Creative Commons Attribution International 4.0 License.}
  \end{minipage}
  \vspace{5pt}
}
\makeatother

\setcopyright{ifaamas}
\acmConference[AAMAS '26]{Proc.\@ of the 25th International Conference
on Autonomous Agents and Multiagent Systems (AAMAS 2026)}{May 25 -- 29, 2026}
{Paphos, Cyprus}{C.~Amato, L.~Dennis, V.~Mascardi, J.~Thangarajah (eds.)}
\copyrightyear{2026}
\acmYear{2026}
\acmDOI{}
\acmPrice{}
\acmISBN{}

\title[Modeling Human Behavior in a Strategic Network Game]{Modeling Human Behavior in a Strategic Network Game with Complex Group Dynamics}

\author{Jonathan Skaggs}
\affiliation{
  \institution{Brigham Young University}
  \city{Provo, UT}
  \country{USA}}
\email{jbskaggs12@gmail.com}

\author{Jacob W.~Crandall}
\affiliation{
  \institution{Brigham Young University}
  \city{Provo, UT}
  \country{USA}}
\email{crandall@cs.byu.edu}

\begin{abstract}
Human networks greatly impact important societal outcomes, including wealth and health inequality, poverty, and bullying.  As such, understanding human networks is critical to learning how to promote favorable societal outcomes.  As a step toward better understanding human networks, we compare and contrast several methods for learning models of human behavior in a strategic network game called the Junior High Game (JHG)~\citep{skaggs2024fostering}.  These modeling methods differ with respect to the assumptions they use to parameterize human behavior (behavior matching vs.~community-aware behavior) and the moments they model (mean vs.~distribution).  Results show that the highest-performing method, called hCAB, models the distribution of human behavior rather than the mean and assumes humans use community-aware behavior rather than behavior matching. When applied to small societies, the hCAB model closely mirrors the population dynamics of human groups (with notable differences).  Additionally, in a user study, human participants had difficulty distinguishing hCAB agents from other humans, thus illustrating that the hCAB model also produces plausible (individual) behavior in this strategic network game.
\end{abstract}


\newcommand{\BibTeX}{\rm B\kern-.05em{\sc i\kern-.025em b}\kern-.08em\TeX}

\begin{document}


\pagestyle{fancy}
\fancyhead{}


\maketitle 


\section{Introduction}
\label{sec:intro}

Human networks (collections of interconnected people) influence how individuals exchange resources~\cite{willer1999network}, form alliances~\citep{cranmer2012toward, gulati1995social}, and navigate competitive pressures~\citep{burt2005brokerage, vu2006learning}. These networks, which are studied in many disciplines~\citep{svendsen2009handbook,levin1998ecosystems}, exhibit emergent behaviors related to inequality~\citep{barabasi1999emergence, salganik2006experimental}, information flow~\citep{Harush2017}, and social behaviors~\citep{shoham1997emergence, christakis2010social, centola2010spread, centola2018behavior, veenstra2021social}. While the study of human networks has yielded deep insights~\citep{newman2003structure,easley2010networks,bar2019dynamics,pagan2019game}, understanding the precise decision-making processes that drive human behavior within these networks remains a major challenge~\citep{young2015evolution}. What determines whether an individual cooperates or competes? How do local interactions scale up to produce global patterns of stability, inequality, or conflict? How might an individual change their behavior to encourage preferred social change?  Addressing these questions requires computational models that capture the complexities of human decision-making in networked environments.

In human networks, individuals make decisions based on personal incentives, past experiences, and the social landscape.  Sometimes individual incentives align with the collective good, like the invisible hand~\citep{smith1776wealth}, but other times they are diametrically opposed, as in the free-rider problem~\citep{olson1965logic}.  These individual actions lead to emergent behaviors that shape societal outcomes.  Extensive studies have explored emergent properties of human networks~\citep{holland1995hidden, jensen2022complexity, shoham2008multiagent}, investigating, for example, how randomness affects network structures~\citep{erdos1959random}, how power distributes among individuals~\citep{barabasi1999emergence}, and how self-organization arises within different social~\citep{watts1998collective} and economic settings~\citep{arthur1994inductive,young2001individual}. These macroscopic perspectives on how human societies function are valuable, yet they are often insufficient for fully understanding human networks and how they evolve.  More detailed models of human decision-making on networks are needed.

To help address this challenge, we model human behavior in a strategic network game designed to capture the emergent properties of complex (networked) societies, including wealth and power distributions and mixing patterns defined by dyadic~\citep{gouldner1960norm, fehr2000fairness}, triadic~\citep{granovetter1973strength, nowak2005evolution}, and group dynamics~\citep{nowak2006five,bowles2011cooperative,henrich2015secret}.  These dynamics emerge due to power asymmetry, mixed-motives, resource management, and interconnectedness~\cite{easley2010networks, jensen2022complexity, thurner2018introduction}, and produce networks that are dynamic, directed, weighted, and signed. Because these dynamics are encoded within the Junior High Game (JHG)~\citep{skaggs2024fostering}, we use this strategic network game to study emergent behavior.

A fundamental challenge in modeling human decision-making in the JHG lies in determining how to parameterize and represent behavior~\citep{rahwan2019machine}.  In this paper, we consider models that make different assumptions about human decision-making. One approach, behavior matching~\citep{axelrod1984evolution}, assumes that agents imitate the observed actions of others, while another approach, community-aware behavior~\citep{skaggs2024fostering}, considers broader structural and social cues within the network.  We also examine how modeling different statistical moments (the mean verses the distribution of behavior) impact model performance.  Together, this produces four different algorithms for modeling human behavior in the JHG.  As a starting point to understanding how well these algorithms model human behavior, we compare and contrast their behavior in small-scale societies.

\begin{figure*}[t]
\centering
    \includegraphics[width=6.8in]{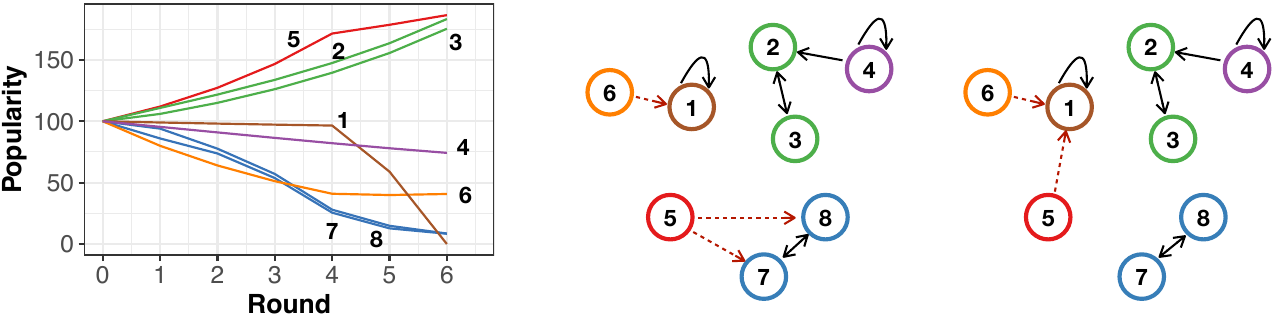}\\ \vspace{.04in}
    {\bf \hspace{1.25cm} (a) \hspace{5.75cm} (b) \hspace{4.7cm} (c)}
    \caption{Example scenario illustrating various JHG dynamics. {\bf (a)} Popularity of the players over time. {\bf (b-c)} Token allocations in rounds 1-4 and 5-6, respectively.  Red dashed arrows indicate tokens used to attack while black solid arrows indicate giving.}
    \label{fig:ejemplo}
    \Description{A figure showing a hypothetical game designed to illustrate various dynamics of the Junior High Game.}
\end{figure*}

We evaluate the ability of these modeling algorithms to produce human behavior in the JHG in two ways.  First, we compare and contrast the {\em population dynamics} of agent societies with those of human groups.  Results show that societies of one of these agents, called hCABs, have population dynamics that mirror (with some differences) human populations. Second, we conduct a user study to identify the extent to which {\em individual} hCABs exhibit human-like tendencies.  Results of this study show that human participants have difficulty distinguishing hCAB agents from other human players.  These findings not only inform the development of models that can simulate human behavior in the JHG, but also offer perspectives on the mechanisms that drive the dynamics of human networks.

\section{A strategic network game}
\label{sec:jhg}

Human networks are punctuated by power asymmetry, mixed motives, resource management, and interconnectedness~\cite{easley2010networks, jensen2022complexity, thurner2018introduction}, which (along with other properties) combine to produce emergent properties of human networks.  These characteristics are modeled in the Junior High Game (JHG)~\cite{skaggs2024fostering}, a strategic game that induces a dynamic (time varying), directed, weighted, and signed network.

We paraphrase \citet{skaggs2024fostering} to review this game: The JHG is played by a set of players $I$ over a series of rounds in which each player $j\in I$ seeks to become {\em popular}.  Initially, every player is assigned a popularity, which changes over time based on interactions, represented by token allocations, among the players.  In every round, each player~$j$ allocates $N$ tokens, each of which it can keep, give to another player, or attack another player.  Keeping tokens positively impacts $j$'s subsequent popularity, while tokens $j$ gives to player $i$ positively impact $i$'s subsequent popularity (but do nothing for $j$).  Finally, when $j$ attacks $i$, it negatively impacts $i$'s subsequent popularity while positively impacting $j$'s.  After all players allocate their tokens in a round, the players' popularities are updated based on the token allocations.  A new round then begins. 

The amount that a token transaction impacts subsequent popularity depends on the popularity of the player allocating the token.  For example, receiving a token from a more popular player results in a higher increase in popularity than receiving a token from a less popular individual.  Thus, players with higher popularity can have higher impact than those with lower popularity.  

To increase understanding of JHG dynamics, consider the example depicted in Figure~\ref{fig:ejemplo}.  First, when players exchange (give) tokens with each other, they rise in popularity (e.g., players~p2 and~p3).  Second, when a player gives tokens without receiving as much in return, they tend to decrease in popularity (e.g., p4).  Third, when a player keeps its token, it maintains much of its popularity, but does not increase in popularity in the absence of other interactions (e.g., p1 in rounds 1-4).  Fourth, attacking those not keeping tokens results in the attacker becoming more popular and the victim becoming less popularity (e.g., p5 attacking p7 and p8 in rounds 1-4).  When other players do not keep or retaliate, attacking is the fastest way to gain popularity.  It also increases the attacker's relative popularity in relation to the victim.  Fifth, keeping tokens blocks attacks.  For example, p6's attacks on p1 in rounds 1-4 are blocked because p1 is keeping its tokens--p1 experiences no loss, while p6 quickly loses popularity due to not gaining from its token allocations.  Finally, attacks that are stronger than the defense (strength is measured as tokens allocated multiplied by player popularity) result in inflicted damage.  For example, p5's and p6's combined attack on p1 in rounds 5-6 is substantially stronger than p1's defense.  Thus, p1 loses popularity and p5 and p6 profit somewhat.

While this contrived example deals primarily with pairwise interactions for illustrative purposes, these dynamics can give rise to complex group behavior and emergent social norms.  Frequently, alliances form and dissolve as players join together to help each other, punish those that violate norms, mitigate powerful enemies, and exploit others.

The complex strategic nature of the JHG gives rise to questions about how humans play it.  Unlike many social dilemmas, players in the JHG must go beyond deciding whether to cooperate or defect.  They must decide with whom they should cooperate (give tokens), when to create new relationships, when to dissolve friendships, when to defend one's self (by keeping tokens), and who they should attack (take tokens) and when.  In the next section, we describe methods that attempt to encode human behavior in the JHG.

\section{Modeling human behavior in the JHG}
\label{sec:modeling}

In this section, we introduce four algorithms that learn models of human behavior in the JHG by observing the play of experienced humans. These algorithms vary based on assumptions encoded in their parameter-based representations and methods for learning parameterizations.  We discuss these design choices in turn.

\subsection{Parameter-based models of behavior}

We could potentially learn models of human behavior in the JHG using a variety of approaches.  For instance, we could take a sort of {\em tabula rasa} approach, in which we feed large amounts of human data into a neural network.  Unfortunately, obtaining large amounts of data of people playing the JHG (or some other strategic network game) is difficult.  Our data set currently contains only a few hundred human player games, and it is unclear how to create human-like data synthetically.  Thus, using a data hungry approach, such as a graph neural network, appears to be somewhat impractical.  Additionally, we would like our model to be interpretable, meaning that it can inform us about the strategies used by human players.  Black-box methods, such as graph neural networks, typically do not provide this desired interpretability.

As such, we consider the ability of parameter-based models to learn human behavior in the JHG.  Such models are based on assumptions of how humans behave, with parameters that modulate behavior.  Formally, let $m$ be the number of parameters in the model.  Then, $\Theta = (\theta_1,\theta_2,\cdots,\theta_m)$ is a parameterization of the model in which $\theta_i$ is the value of the $i^{\rm th}$ parameter.

In this paper, we consider two models: (1) Behavior Matching (or tit-for-tat; TFT)~\cite{axelrod1984evolution} and (2) Community-Aware Behavior (CAB)~\cite{skaggs2024fostering}.

\textbf{Behavior Matching (TFT)}:  In this model, a player reciprocates the actions that other players directed towards them in the previous round \cite{axelrod1984evolution}.  We consider this model due to its prevalence in prior work.  TFT and its variants have repeatedly shown to be effective and robust in repeated prisoner's dilemmas (e.g., \cite{axelrod1980effective, nowak1992tit, grant2013give,nowak1992evolutionary}), a common domain for studying when and how humans and other agents cooperate with each other.  Furthermore, TFT and its variants are frequently used to describe (successful) human behavior (e.g., \cite{axelrod1984evolution,grant2013give}).  As such, we consider whether human players use TFT-like strategies in the JHG, and, thus, whether a parameterized TFT algorithm can be tuned to successfully mimic human behavior.

However, because a player's resources are limited in the JHG, defining TFT in this domain is not straightforward, as a player may receive more or less in a round than it has the ability to reciprocate.  In such cases, an agent must determine what to do with extra tokens (when receiving too much) or who to exclude (when not receiving enough).  We have defined a TFT algorithm for the JHG whose behavior is tuned with seven parameters that determine how the agent allocates tokens in the first round, what it seeks to match (token allocations or influence), and how it deals with excess and shortfalls. Details are given in the supplementary material (SM-2).

\textbf{Community-Aware Behavior (CAB)}: The second parameter-based model we consider is the CAB ({\em C}ommunity {\em A}ware {\em B}ehavior) algorithm~\citep{skaggs2024fostering}.  While using behavior matching to model humans assumes that humans cooperate and attack each other due to reciprocation, using the CAB algorithm assumes that humans help and attack each other based on group dynamics~\cite{moffett2019human}.  Players in the same group support (give to) each other, work together to exploit other groups, and defend each other against the attacks of outsiders.  CAB agents follow a parameterized algorithm for how humans form, join, and modify these groups over time.

A CAB agent determines token allocations in a round using a three-step process.  First, it assigns players to groups based on past token allocations.  Second, it chooses a (potentially new) group it would like to form, support, or join.  Finally, it selects token allocations designed to build up its selected group, establish itself in the group, and deal with out-of-group threats.  CAB behavior is defined by thirty parameters which determine how it performs each of these steps.  We use the CAB implementation provided by~\citet{skaggs2024fostering}.

Assuming that human behavior in the JHG can be modeled using either of these parameterized algorithms (TFT or CAB) is most certainly incorrect at some level.  However, identifying the extent to which these models can and cannot approximate human behavior can shed valuable insights into human networks and how human behavior can be effectively simulated in them.

\subsection{Learning Parameterizations}

To determine how well TFT and CAB parameter models can model human behavior in the JHG, we learn parameterizations of each model from human data using two different methods: Particle Swarm Optimization (PSO)~\cite{kennedy1995particle} and Evolutionary Population Distribution Modeling (EPDM), a methodology proposed in this paper.  Both algorithms take as input a parameter-based model $\mathcal{M}$ and output one or more parameterizations of that model.  

\textbf{Particle Swarm Optimization (PSO)}:  In PSO, a set of particles (i.e., parameterizations) are first randomly generated.  Over time, these particles are adapted in a way that both explores the parameter space and converges toward the best parameterization, or the parameterization that minimizing the error, over all data samples, between the token allocations made by human players in the data and the token allocations made by the model (given the parameterization specified by the particle) in the same situations.  PSO produces a parameterization designed to describe {\em mean} human behavior represented in the data.

We experimented with two error (or distance) functions: (1) mean squared error and (2) a custom built function which measures how well a token allocation profile matches a target profile across a variety of properties.  Because the second function outperformed the first in all cases, we used it in our experiments.  This function scores the degree to which token allocation profile $\mathbf{b}$ matches the target profile $\mathbf{a}$ as follows:
\begin{equation}
{\mathcal S}(\mathbf{a},\mathbf{b}) = {\mathcal S}^+(\mathbf{a},\mathbf{b}) + {\mathcal S}^-(\mathbf{a},\mathbf{b}) + {\mathcal S}^{\rm keep}(\mathbf{a},\mathbf{b}) - {\mathcal P}(\mathbf{a},\mathbf{b}).
    \label{eq:allocation_score}
\end{equation}
Here, ${\mathcal S}^+(\mathbf{a},\mathbf{b})$ scores the degree to which $\mathbf{b}$ (a)~has positive allocations to the same number of players as $\mathbf{a}$, (b) has positive allocations to the same players as $\mathbf{a}$, (c) matches the total number of tokens allocated for giving as $\mathbf{a}$, and (d) gives the same number of tokens to each player as $\mathbf{a}$.  ${\mathcal S}^-(\mathbf{a},\mathbf{b})$ scores the degree to which $\mathbf{b}$ (a)~has a negative allocation if $\mathbf{a}$ has a negative allocation, (b)~has the same number of tokens used for taking as $\mathbf{a}$, and (c)~takes tokens from the same players as $\mathbf{a}$.  ${\mathcal S}^{\rm keep}(\mathbf{a},\mathbf{b})$ compares the number of tokens kept in allocations $\mathbf{a}$ and $\mathbf{b}$.  Finally, ${\mathcal P}(\mathbf{a},\mathbf{b})$ is a penalty term when allocation $\mathbf{b}$ gives to some player~$i$ but allocation $\mathbf{a}$ takes from player~$i$ or vice versa.  Details related to ${\mathcal S}(\mathbf{a},\mathbf{b})$ are given in SM-4.

\textbf{Evolutionary Population Distribution Modeling (EPDM)}:
Because mean human behavior, when followed by all individuals, may not produce human-like population dynamics or realistic individual behavior, we introduce a second method, called EPDM, for modeling the {\em distribution} of human behavior. The EPDM algorithm, outlined in Algorithm~\ref{alg:epdm}, receives as input a parameter-based behavior model $\mathcal{M}$ (we consider TFT and CAB in this paper) and a set of player games $\Gamma$ (from the training data).  Each player game $\gamma \in \Gamma$ specifies the token allocations made by a human player in each round $t$ of a game, along with the game state available to the player at the beginning of that round.  This game state consists of a vector ${\mathcal P}(t)$ specifying the current popularity of each player, a matrix ${\mathcal I}(t)$ specifying the influence that the players have on each other, and a matrix $X(t)$ specifying the token allocations made by all players in that round.  The algorithm outputs a set of $N$ parameterizations of $\mathcal{M}$ that best estimate the distribution of player strategies in the player games $\Gamma$.  We used $N=100$.

\begin{algorithm}[tb]
   \caption{{\em EPDM} -- Computes $N$ parameterizations to estimate a distribution of strategies}
   \label{alg:epdm}
\begin{algorithmic}
   \STATE {\bfseries Input:} $\mathcal{M}$ (parameter-based model), $\Gamma$ (player games), and $G$ (\# generations)
   \STATE {\bfseries Initialize}: Set $g = 0$ and randomly generate $\Pi(0)$
   \WHILE{$g < G$}
   \STATE (1) $\forall \pi \in \Pi(g)$, compute $\Gamma^\pi(g)$ (player games in which $\pi$ is a top performer)
   \STATE (2) $\Phi(g) =$ FindCoreSet$(\Gamma, \Pi(g), \forall {\pi \in \Pi(g)},~ \Gamma^{\pi}(g))$
   \STATE (3) Construct $\Pi(g+1)$:
   \STATE ~~~~~~- $\Pi(g+1) = toList(\Phi(g))$
   \STATE ~~~~~~- Add $N-|\Phi(g)|$ additional parameterizations to $\Pi(g+1)$ using genetic evolution on $\Pi(g)$
   \STATE (4) $g = g + 1$
   \ENDWHILE
   \STATE $\Pi(G) =$ Resample $\Pi(G-1)$ proportional to fitness
   \STATE {\bf return} $\Pi(G)$
\end{algorithmic}
\end{algorithm}

EPDM learns this distribution of parameterizations using genetic evolution.  Initially, a set of $N$ parameterizations (denoted $\Pi(0)$) are randomly generated.  Over a series of $G$ generations, the pool of parameterizations is evolved to better estimate the distribution of strategies used by human players in the supplied player games.  Each generation $g$ proceeds in three steps.  In step 1, the set of player games for which parameterization $\pi \in \Pi(g)$ is a top performer is computed.  This set, denote $\Gamma^\pi(g)$, is determined as follows.  Let $\mathbf{a}^\gamma(t)$ denote the token allocations made in player game $\gamma \in \Gamma$ in round $t$.  Also, let $\mathbf{a}^\pi(t)$ be the allocations that parameterization $\pi$ would make in round $t$ given the game history (prior to round $t$).  Then, the average error of $\pi$ in player game $\gamma$ is
\begin{equation}
    \delta^\pi_\gamma = \frac{1}{T_\gamma} \sum_{t=1}^{T_\gamma} dist(\mathbf{a}^\gamma(t), \mathbf{a}^\pi(t)),
    \label{eq:error_round}
\end{equation}
where $T_\gamma$ is the number of rounds in player game $\gamma$ and $dist(\mathbf{a}, \mathbf{b})$ is the error function specifying the distance between allocations $\mathbf{a}$ and $\mathbf{b}$ (we use $-{\mathcal S}(\mathbf{a},\mathbf{b})$; Eq.~\ref{eq:allocation_score}).  $\pi$ is a top-performing parameterization for player game $\gamma$ if $\delta^\pi_\gamma - \min_{\pi^\prime \in \Pi} \delta^{\pi^\prime}_\gamma < \varepsilon$.  We use $\varepsilon = 0.01$.

In step 2 of Algorithm~\ref{alg:epdm}, a core set of parameterizations $\Phi(g) \subseteq \Pi(g)$ is computed using the greedy process defined in the function {\em FindCoreSet} (see Algorithm~\ref{alg:find_core_set}).  {\em FindCoreSet} repeatedly adds the parameterization in $\Pi(g)$ that is a top-performer in the most player-game sets that have not yet been represented until all player games are represented by at least one top-performing parameterization.

\begin{algorithm}[tb]
   \caption{{\em FindCoreSet} -- Selects a set of parameterizations from the current set of parameterizations $\Pi(g)$ to describe the set of observed human strategies.}
   \label{alg:find_core_set}
\begin{algorithmic}
   \STATE {\bfseries Input:} 
   \STATE ~~~- $\Gamma$ (set of player games)
   \STATE ~~~- $\Pi(g)$ (list of parameterizations used in generation $g$)
   \STATE ~~~- $\Gamma^\pi(g)$ for each $\pi \in \Pi(g)$ (player games for which $\pi$ is a top-performing parameterization)
   \STATE {\bfseries Initialize}: 
   \STATE ~~~- $\hat{\Gamma} = \Gamma$
   \STATE ~~~- $\hat{\Gamma}^\pi(g) = \Gamma^\pi(g)$
   \STATE ~~~- $\Phi = \emptyset$
   \WHILE{$\hat{\Gamma} \neq \emptyset$}
   \STATE $\pi^* = \max_{\pi \in \Pi(g) \setminus \Phi} |\hat{\Gamma}^\pi(g)|$
   \STATE $\hat{\Gamma} = \hat{\Gamma} \setminus \hat{\Gamma}^{\pi^*}$
   \STATE $\Phi = \Phi \cup \{\pi^*\}$
   \STATE $\forall \pi \in \Pi(g), \hat{\Gamma}^\pi(g) = \hat{\Gamma}^\pi(g) \setminus \hat{\Gamma}^{\pi^*}(g)$
   \ENDWHILE
   \STATE {\bf return} $\Phi$
\end{algorithmic}
\end{algorithm}

Finally, in step 3, the EPDM algorithm constructs the set of parameterizations used in the subsequent generation.  First, each parameterization from the core set $\Phi(g)$ is added to $\Pi(g+1)$.  The remaining parameterizations are then generated using genetic evolution on the parameterizations in $\Pi(g)$ using selection, crossover, and mutation.  First, two parameterizations are randomly (based on fitness, defined as the number of player games for which the parameterization is a top performer) selected from $\Pi(g)$.  That is, the probability parameterization $\pi$ is selected is $\frac{|\Gamma^\pi(g)|}{\sum_{\pi'} |\Gamma^{\pi'}(g)|}$, where $|\Gamma^\pi(g)|$ is the cardinality of the set $\Gamma^\pi(g)$.  Second, a new parameterization is formed from the two parameterizations selected in the prior step.  In this parameterization, the $i^{\rm th}$ value is determined in one of three ways.  With probability 0.03, it is a random value in the range $[0,100]$.  Otherwise, the value is chosen to be the $i^{\rm th}$ value of one of the two selected parameterizations (chosen randomly).  With probability 0.12, this value is then randomly shifted up or down 0 to 5 values (truncated to be between 0 and 100, inclusive).

After all generations are completed, the last step of the EPDM algorithm (Algorithm~\ref{alg:epdm}) consists of resampling the parameterizations evaluated in the last generation $\Pi(G-1)$ based on fitness.  This is done by randomly selecting $N$ parameterizations from the list $\Pi(G-1)$.  In selecting each parameterization, the probability that parameterization $\pi$ is chosen is given by $\frac{|\Gamma^\pi(g)|}{\sum_{\pi'} |\Gamma^{\pi'}(g)|}$.  Note that a parameterization can be chosen more than once.


\begin{table*}[t]
    \caption{Metrics used to compare the population dynamics of human and agent societies.}
    \label{tab:metrics}
    \centering
    \begin{tabular}{p{0.14\linewidth}p{0.47\linewidth}p{0.32\linewidth}} \hline
        \textbf{Category} & \textbf{Description} & \textbf{Metrics} \\ \midrule
        Wealth \& Power & Wealth and influence of society members & Mean Popularity, Gini Index \\ 
        Economic Behavior & High-level actions (give, take, keep) taken by individuals in society & \% Give, \% Take, \% Keep, Evolution Coefficient \\ 
        Mixing Patterns & Connections across society, including dyadic, triadic, and group relationships & Reciprocity, Density, Entropy, Polarization \\ 
        Summary & Comparison to humans across all metrics & Mahalanobis distance \\ \hline
    \end{tabular}
\end{table*}

\section{Evaluating population dynamics} 
\label{sec:analyticalresults}

Our goal is to model human behavior in the JHG so as to satisfy two criteria.  First, the population dynamics of agent collectives should mirror the population dynamics of human societies.  Second, individual agents should, ideally, also exhibit human-like behavior.  In this section, we evaluate four models, created by combining parameter-based models (hTFT and hCAB\footnote{hCAB and hTFT denote parameterizations of CAB and TFT models, respectively, learned from {\em h}uman data.}) with modeling algorithms (PSO and EPDM), with respect to the first criteria.  In Section~\ref{sec:userstudyresults}, we address the second criteria.  

We note that human behavior is likely impacted by group size.  Understanding human behavior given all group sizes is valuable.  However, given that our data sets (Section~\ref{sec:datasets}) consist of small-scale human groups (5-12 individuals), we only consider groups of this size in the evaluations made in this paper.  While we anticipate that human behavior in larger groups could be modeled with the same algorithmic mechanisms, we leave such evaluations to future work.

\subsection{Experiment Design}

We simulate societies of hTFT-PSO, hTFT-EPDM, hCAB-PSO, and hCAB-EPDM agents in the JHG, and compare their dynamics to those of games played by human players.  We also consider two baseline agents: (1) Random, which randomly allocates tokens with the percentages of give, take, and keep that match the percentages observed in the training set (Section~\ref{sec:datasets}) and (2)~eCAB agents~\citep{skaggs2024fostering}.  eCABs are CAB agents with parameterizations learned via evolutionary simulations rather than from human data.  eCABs display somewhat effective behavior in the JHG, though their strategies are distinct from those used by humans~\cite{skaggs2024fostering}. 

To evaluate these agents, we compare their population dynamics with those of human players observed in the 15 games in our test set (Section~\ref{sec:datasets}).  Simulations for agent populations were conducted with identical settings (initial popularities, number of agents, and game length) as games found in the test set.  For each game in the test set, four simulations of each agent population were conducted with identical numbers of agents and rounds, bringing the total number of simulations for each agent to 60. In setting up experiments in this way, we can directly compare human and agent societies.

For simulations with hTFT-EPDM, hCAB-EPDM, and eCAB agents, parameterizations were randomly selected from the learned pool of strategies for each game.  On the other hand, for hTFT-PSO and hCAB-PSO, the parameterizations to which the PSO algorithm converged in four separate runs of the PSO algorithm were used.

\subsection{Data Sets}
\label{sec:datasets}

Our data set, which is provided in the supplementary material, consists of 66 games played primarily by experienced human players.  47 of these 66 games had only human players.  The other 19 games were played by both human and bot players.  Of the 47 all-human games, we randomly selected 15 of these games for the {\em test set}.  The remaining 51 games were designated as {\em training data}.  Games in the training set had 5-12 players with an average of 8.08 players (of which 6.57 were humans).  Across these games, we have a total of 335 human {\em player games} (which comprise $\Gamma$).  The average number of rounds in these player games was 24.96.  Games in the test set were played by 6-11 human players (average was 7.93 players).  Games with more than 30 rounds were truncated at 30 rounds.

\subsection{Metrics to Evaluate Population Dynamics}
\label{sec:metrics}

We evaluate population dynamics using the metrics summarized in Table~\ref{tab:metrics}.  These metrics are grouped into four categories, each designed to evaluate a particular aspect of population dynamics.  Metrics in the first category quantify the {\em wealth and power} wielded by society members.  We use Mean Popularity, an approximation of Katz centrality~\citep{katz1953new}, and Gini Index, which measures the distribution of wealth across society.  {\em Economic behavior} summarizes player actions across society, including the percentages of token allocations of each type (keep, take, and give) and the average change in player token allocations from round to round (Evolution Coefficient).  Next, we assess societal {\em mixing patterns}, or the way that players both join together in pairs and groups, and fragment and segregate.  These mixing patterns are quantified using two measures of reciprocity, density, entropy, and polarization.  Finally, we summarize the dissimilarity between agent and human populations across all these metrics using Mahalanobis distance~\citep{mahalanobis1936generalized}.  Formal definitions of each metric, as well as motivation for why we chose these (as opposed to other) metrics, are provided in SM-5.

\subsection{Results}

\begin{figure*}[t]
    \begin{center}
    \includegraphics[width=2.21in]{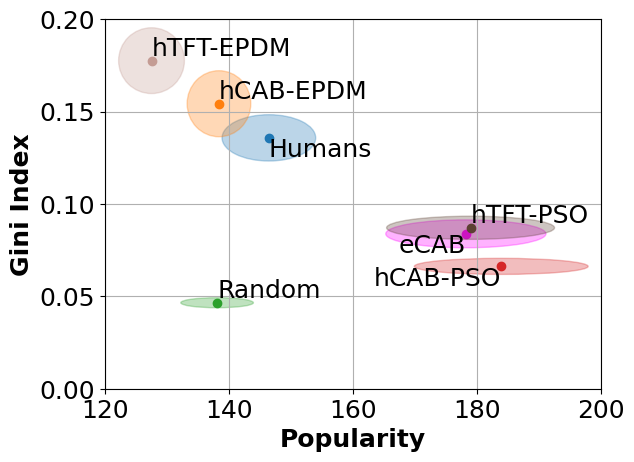} \hspace{.1in}
    \includegraphics[width=2.10in]{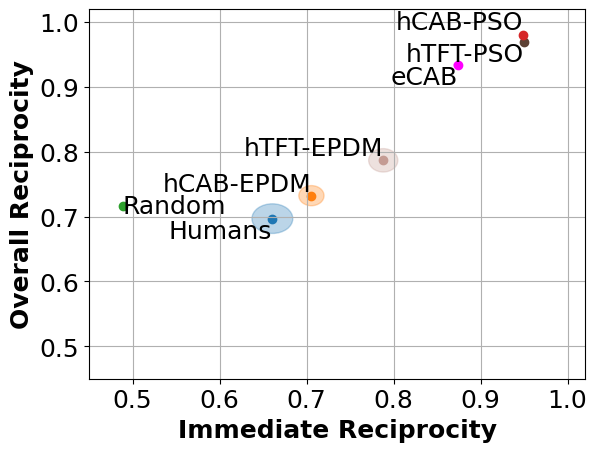} \hspace{.1in}
    \includegraphics[width=2.21in]{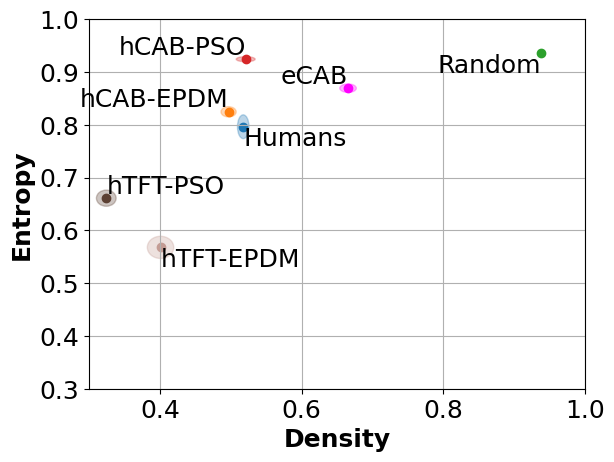} 
    \end{center}\vspace{0in}
    \begin{center}
    {\bf (a)} \hspace{2.1in} {\bf (b)} \hspace{2.1in} {\bf (c)} 
    \end{center}
    ~~\vspace{.1in} \\ 
    \hspace{-1.25in}
    \renewcommand{\tabcolsep}{1pt}
    {\small \begin{tabular}{lccc}
         \hline
        \textbf{Society}   & \textbf{\% Give} & \textbf{\% Keep} & \textbf{\% Take} \\
        \midrule
        Humans           & $74.9\pm1.0$   & $18.8\pm0.7$   & $6.2\pm0.8$   \\
        Random           & $72.0\pm0.0$   & $24.0\pm0.0$   & $4.0\pm0.0$   \\
        hTFT-PSO   & $97.3\pm1.8$   & $2.7\pm1.8$    & $0.0\pm0.0$   \\
        hTFT-EPDM    & $69.9\pm0.3$   & $14.9\pm0.8$   & $15.2\pm0.4$  \\
        eCAB             & $94.7\pm1.8$   & $5.2\pm0.0$    & $0.0\pm0.0$   \\
        hCAB-PSO     & $100.0\pm0.0$  & $0.0\pm0.0$    & $0.0\pm0.0$   \\
        hCAB-EPDM    & $73.6\pm1.0$   & $19.1\pm0.4$   & $7.3\pm0.6$   \\
        \hline
    \end{tabular}} \hspace*{3.7in}
    \\ \vspace{-1.3in}~\\ \hspace*{2.2in}
    \includegraphics[width=4.7in]{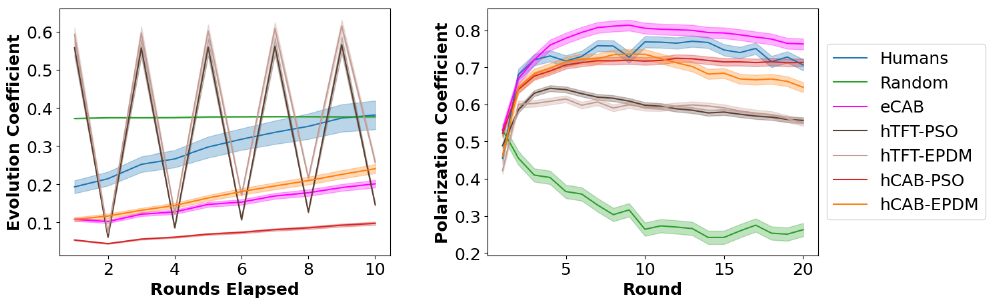}
    \\ 
    {\bf (d)} \hspace{2.1in} {\bf (e)} \hspace{2.1in} {\bf (f)}     
    \caption{Comparisons of human and agent societies. Dots (a-c), numbers (d), and lines (e-f) denote the mean, error ovals (which do not account for covariance), $\pm$, and error ribbons show the standard error.}
    \label{fig:los-resultados}
    \Description{A comparison of population dynamics of various agent populations with human populations across many different metrics, including popularity and Gini index, reciprocity, polarization, density, entropy, and economic action.}
\end{figure*}

Figure~\ref{fig:los-resultados} displays measures of human and agent population dynamics with respect to individual metrics.  We discuss each by category.

\textbf{Wealth and Power}:
Figure~\ref{fig:los-resultados}a compares agent and human populations with respect to Mean Popularity and Gini Index.  In these two regards, hCAB-EPDM populations most closely mirror those of humans.  Random populations tend to have both low Gini Index and low mean popularity.  hTFT-EPDM populations tend to have slightly higher Gini Index and lower popularity, while hTFT-PSO, hCAB-PSO, and eCAB populations all have higher Mean Popularity and lower Gini Index.  While high Mean Popularity and lower Gini Index is desirable (though this does not necessarily indicate robust behavior in general), our objective is to identify models that mimic human behavior.  With respect to Mean Popularity and Gini Index, the CAB model coupled with EPDM parameter modeling results in the model that most closely reflects human population dynamics.

\textbf{Economic Behavior}: 
In the JHG, wealth outcomes are the result of economic behavior.  eCAB agents and algorithms that model parameters using PSO all have higher Mean Popularity and lower Gini Index than human populations.  These agents allocate most of their tokens to giving.  They keep and take little, if any (Figure~\ref{fig:los-resultados}d).  On the other hand, humans and agents modeled with EPDM have higher variability in strategies across players and tend to keep and take more tokens.  This produces greater societal inequality and less overall prosperity.  As such, agents produced using EPDM have token allocation distributions that closer to those of humans.

The Evolution Coefficient measures the consistency of player token allocations over rounds.  Figure~\ref{fig:los-resultados}e shows that none of the agent populations display similar behavior to humans in this regard.  CAB agents are too consistent in their allocations (though hCABs are less consistent than eCABs), showing less change in token allocations from round to round than humans.  On the other hand, TFT agents display a sort of multiple personality disorder.  The act of reciprocating actions from the previous round produces oscillating changes in token allocations that are clearly not human-like.  However, the differences in token allocations of TFT agents over two rounds are quite small.  

\textbf{Mixing Patterns}: We measure two forms of reciprocity: Immediate Reciprocity (reciprocation from one round to the next) and Overall Reciprocity (reciprocation over a complete game).  In the JHG, humans have moderate levels of both forms of reciprocity (Figure~\ref{fig:los-resultados}b).  EPDM agents exhibit similar levels of reciprocity, with hCAB-EPDM's rates being only slightly higher than people's.  On the other hand, eCABs and PSO reciprocation rates are far higher than those of the humans in evaluated in our test set.

The density and entropy measures, shown in Figure~\ref{fig:los-resultados}c, provide information about the number (as measured by Density) and strength (as measured by Entropy--this metric measures whether all trading partners are valued similarly) of trading partners.  In these regards, hCAB-EPDM agents once again tend to be very similar to those of human players.  hCAB-PSO agents have similar Density, but have higher Entropy.  On the other hand, agents modeled using the TFT model tend to be less similar to humans in both regards.  Thus, hCAB agents mirror these human behaviors better than TFT agents, with hCAB-EPDM being the closest.

Finally, we consider Polarization (Figure~\ref{fig:los-resultados}f), which describes the degree to which players segregate into groups.  Since the CAB algorithm is based on group formation, while TFT is not, it is not surprising to observe that CAB agents display higher Polarization than TFT agents.  These higher levels are similar to the Polarization measures we observe in human games in the test set.  This again shows that the CAB model can be tuned to better model human dynamics than the TFT (matching) model.

\begin{table}[t]
    \caption{Mahalanobis distance (lower is better) between human and agent populations over all metrics.  $p>0.05$ indicates that we cannot reject the null hypothesis (that they come from the same distribution).} 
    \label{tab:mahalanobis}
    \centering
    \renewcommand{\tabcolsep}{2.7pt}
    \begin{tabular}{lcc} \hline
        \textbf{Algorithm} & {\bf Mahalanobis Distance} & {\bf Chi-Square Test} \\
        \midrule
        Random & 18.822 & $p < 0.001$ \\
        eCAB & ~~7.089 & $p < 0.001$ \\
        hTFT-PSO & 11.565 & $p < 0.001$ \\
        hTFT-EPDM & ~~7.656 & $p < 0.001$ \\
        hCAB-POS & ~~9.216 & $p < 0.001$ \\
        {\bf hCAB-EPDM} & {\bf ~~3.393} & $\mathbf{p = 0.486}$ \\ 
        \hline
    \end{tabular} \vspace{.05in}
\end{table}

\textbf{Summary Metric}: Table~\ref{tab:mahalanobis} gives the Mahalanobis distance between the various agent populations and our human population across the eleven individual metrics.  Of the six agent populations, hCAB-EPDM populations most closely mirror human societies from our test set.  Indeed, a Chi-Square test comparing human and hCAB-EPDM populations produces a p-value ($p = 0.486$) indicating that we cannot reject the null hypothesis (that they come from the same distribution), thus indicating a somewhat close fit across these measures.  All other agent populations were statistically distinct from human populations ($p < 0.001$ in all cases).

These results show that the CAB algorithm paired with EPDM parameterization modeling produces agent populations whose dynamics are surprisingly close (with some variation--notably the Evolution Coefficient) to those of human populations in the JHG (first criteria).  Is this relatively similar behavior limited to population dynamics, or does it also carry over into individual behaviors?  In the next section, we complement the quantitative comparisons of human and agent populations with a user study designed to evaluate whether individual hCAB-EPDM agents plausibly exhibit human behavior (second criteria).

\section{Interacting with human players}
\label{sec:userstudyresults}

To further illustrate the behavior of hCAB-EPDM agents (hereafter referred to as hCABs), we conducted a user study in which experienced human players interacted with hCABs in the JHG.  From this study, we (1)~illustrate the behavior of individual hCABs when interacting with humans and (2)~quantify how well human participants were able to distinguish hCABs from humans.

\subsection{User Study Design}

We recruited eight people, each of which had prior experience playing the JHG, to participate in the study.  The study proceeded in a series of four periods.  In each period, the participants were randomly divided in half.  Each group of four participants was paired with four (randomly selected) hCABs to form two games played by eight players each.  Thus, eight games were played across the four periods.  Games lasted for 21, 20, 17, and 15 rounds in Periods 1-4, respectively (players were not told game lengths).  For consistency, only the first 15 rounds of each game were used in our data analysis.  

Participants were asked to play so as to become as popular as possible by the end of each game, while also identifying whether associates were humans or bots.  Participants were told in advance that each game was composed of four people and four bots.  After each game, participants completed a survey asking them to identify each associate as a human or a bot.  We also compare humans and hCABs with respect to their popularity and the percentage of tokens used to give, keep, and take.

\subsection{Results}

\begin{figure}
\includegraphics[width=3.2in]{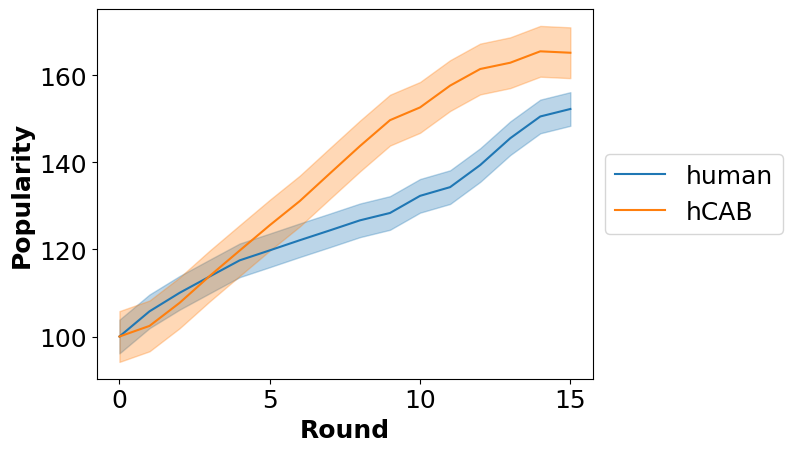}
\caption{Mean popularity over time of humans and hCABs in the user study.  Error ribbons show the standard error.}
\Description{A comparison of the mean popularity over time of humans and bots in the user study.}
\label{fig:userstudy_meanpop}
\end{figure}

\begin{table}
    \caption{The percentage of tokens allocated to giving, keeping, and taking by humans and hCABs in the user study.}
    \label{tab:userstudy_tokens}
    \centering
    \begin{tabular}{lccc} \hline
        \textbf{Society}   & \textbf{\% Give} & \textbf{\% Keep} & \textbf{\% Take} \\
        \midrule
        Humans           & ~~60   & 25   & 15   \\
        hCABs           & ~~75   & 20   & ~~5   \\
        \hline
    \end{tabular} 
\end{table}

\begin{table}
    \caption{The percentage of players correctly identified as human or bot by human participants verses random chance.}
    \label{tab:userstudy_guesses}
    \centering
    \begin{tabular}{lc} \hline
        \textbf{Guesser}   & \textbf{\% Correct Designation} \\
        \midrule
        Humans           & 56.7     \\
        Chance           & 57.1     \\
        \hline
    \end{tabular}
\end{table}



\begin{figure*}[ht]
    \centering
    \includegraphics[width=6.5in]{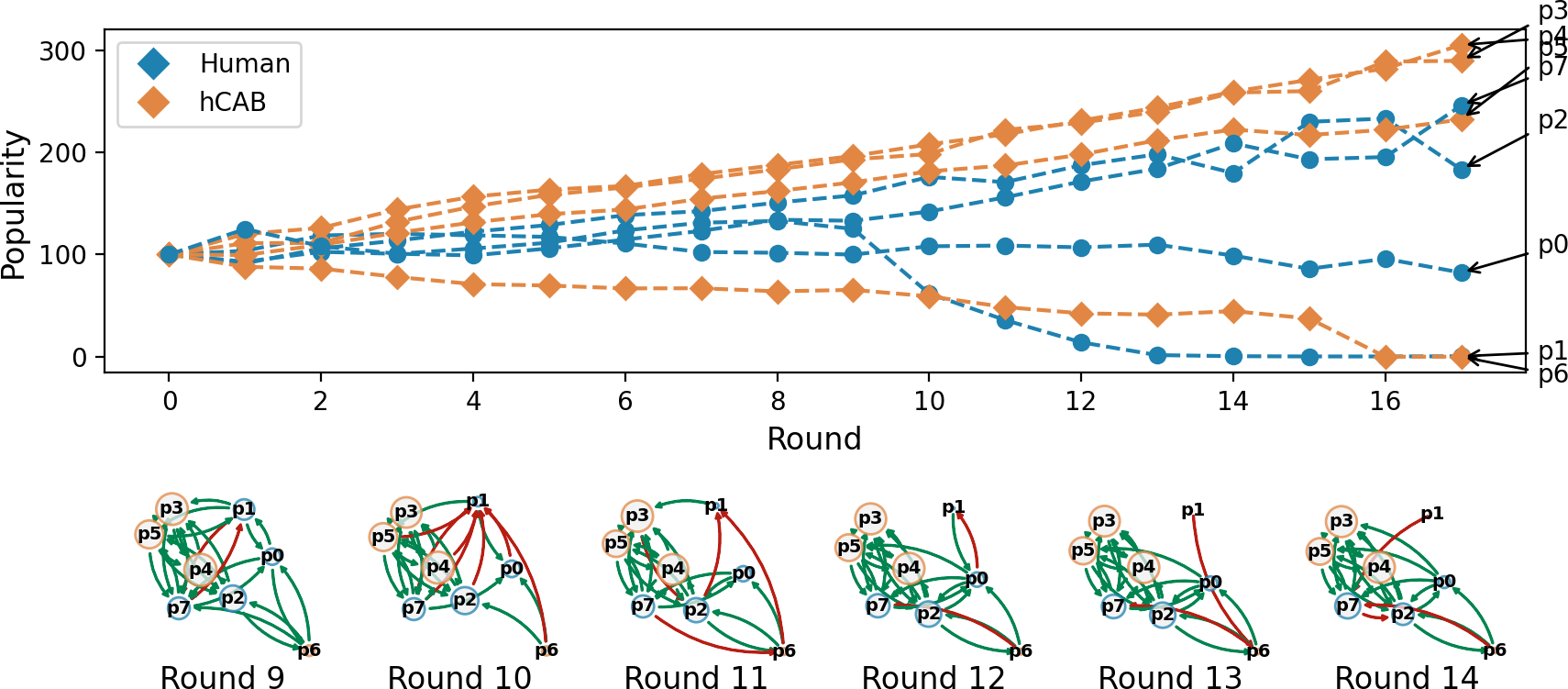}
    \caption{Illustrative game showing the popularity of players over time (top) and network structure in select rounds (bottom).  In the graphs, arrows indicate token transactions in the current round (green = give; red = take), while nodes more connected historically tend to be closer to each other.}
    \label{fig:illustrative_example}
    \Description{Illustration of a human-bot game showing popularity of the players over time as well as relationship graphs in select rounds.}
\end{figure*}

Over the eight games, hCABs had slightly higher average popularity than humans in rounds 5-15 (Figure~\ref{fig:userstudy_meanpop}).  This was potentially caused by some human participants being quite aggressive, attacking substantially more than hCABs (Table~\ref{tab:userstudy_tokens}).  As attacks sometimes lead to retaliation and prolonged fighting, these players often subsequently lost popularity.  While the distribution of human actions was different than humans in our data sets (see previous section), average hCAB behavior remained about the same.

Despite these differences between these particular human participants and hCABs, participants were unable to distinguish each other from hCABs on average.  Table~\ref{tab:userstudy_guesses} shows that participants correctly identified whether their associates were human or bot 56.7\% of the time.  Given that participants knew that four of their associates were bots and three were humans, random chance would have produced a correct detection rate of 57.1\%. Therefore, these participants were, on average, no better than random in identifying whether their associates were humans or hCABs.

Viewing hCAB behavior in individual games provides further evidence of this claim.  For example, consider Figure~\ref{fig:illustrative_example}, which depicts one of the eight games from the study (similar graphs for the other games are provided in SM-7).  In Round 9, two human players (p1 and p2) attacked each other.  Interestingly, three of the four hCABs and the two other humans all joined with p7 in attacking p1 in Round 10 (which greatly weakened p1 for the remainder of the game).  That the hCABs all chose to take the same side in the {\em dispute} between p7 and p1 as the other human players (siding with the better connected individual) illustrates the ability of hCABs to reason about groups and power in a way that led them to mirror human behavior in this situation.  Similar observations can be made about how hCABs joined together in subgroups with humans and other hCABs in these games.

These results give credence to the argument that the behaviors of individual hCABs do seem human-like.  Similarities between humans and hCABs can be seen in both population dynamics and in individual behaviors.  However, we caution against over-extrapolating these results.  The user study measured only a small set of experienced (eight) human players.  Differences could emerge in a larger user study.  Furthermore, population dynamics (see the previous section) do show differences between humans and hCABs.  That said, these results do, by and large, indicate that hCABs are surprisingly difficult to distinguish from humans.

\section{Conclusions and delimitations}
\label{sec:conclusion}

In this paper, we compared and contrasted several algorithms for modeling human behavior in a strategic network game called the Junior High Game (JHG).  We considered approaches that differ with respect to the underlying assumptions of human behavior they use (behavior matching verses community-aware behavior) and the statistical moments they model (mean verses distribution).  Results indicate that a parameter-based community-aware model (CAB~\citep{skaggs2024fostering}) coupled with evolutionary population distribution modeling (EPDM) can produce agents that closely reflect human population dynamics in the JHG (albeit with differences).  Furthermore, results of a user study indicate that the individual behaviors of these agents are difficult to distinguish from those of humans.

The CAB model consists of a variety of hand-coded reasoning functions modulated by thirty parameters.  Such a modeling approach has advantages and disadvantages compared to a more {\em tabula rasa} approach (such as a graph neural network).  One advantage is that an effective model of human behavior can be derived using only a small data set.  This is valuable, since obtaining large amounts of quality data in strategic network games can be challenging.  A second advantage is that the trained strategies are interpretable, thus potentially shedding insight on human behavior.  On the other hand, a tabula rasa approach could potentially learn other kinds of strategies, and is therefore a potentially promising area for future work once larger data sets are obtained.  hCAB-EPDM provides a baseline for evaluating such future methods.  

While our results indicate that hCAB-EPDM agents are, perhaps surprisingly, effective models of human behavior in the JHG, we caution against overly broad extrapolations.  For example, the groups modeled in this paper were small (6-11 players).  Larger societies have different characteristics.  Additionally, the hCAB-EPDM agents used in this paper were trained on a small data set.  Given the complex state and strategy space of the JHG, other human players (from, for example, different cultures) could potentially use different strategies not found in this data set.  Though the CAB algorithm can be configured (via its parameters) to model many different kinds of behavior, it is possible that other communities of human players could display vastly different norms of acceptable behavior that may not be effectively modeled by parameterizations of the CAB algorithm.  Finally, the value of the results described in this paper somewhat hinges on the degree to which the strategic network game we selected (the JHG) models human networks.  While the JHG models important properties of such networks, the world is certainly more complex than any game can model.  Future work should consider refinements and enhancements to study other complexities of human networks.  These and other future works can potentially help us to continue to better understand and model human networks.

\section{Supplementary Material (SM)}

Supplementary documentation, results, and code are available at:
\url{https://github.com/jakecrandall/AAMAS2026-hCABs.git}



\bibliographystyle{ACM-Reference-Format} 
\bibliography{bib}


\end{document}